# Highly Linear, Broadband Optical Modulator Based on Electro-optic Polymer

Xingyu Zhang,[1] Beomsuk Lee,[1] Che-yun Lin,[1] Alan X. Wang,[2] Amir Hosseini,[3] and Ray T. Chen[1]

[1]The University of Texas at Austin, 10100 Burnet Rd, MER160, Austin, TX, 78758, USA,
[2]Oregon State University, 3097 Kelley Engineering Center, Corvallis, OR, 97331, USA
[3]Omega Optics, Inc., 10306 Sausalito Drive, Austin, TX, 78759, USA

**Abstract:** In this paper, we present the design, fabrication and characterization of a traveling wave directional coupler modulator based on electro-optic polymer, which is able to provide both high linearity and broad bandwidth. The high linearity is realized by introducing $\Delta\beta$-reversal technique in the two-domain directional coupler. A travelling wave electrode is designed to function with bandwidth-length product of 302GHz•cm, by achieving low microwave loss, excellent impedance matching and velocity matching, as well as smooth electric field profile transformation. The 3-dB bandwidth of the device is measured to be 10GHz. The spurious free dynamic range of 110dB$\cdot$3Hz$^{2/3}$ is measured over the modulation frequency range 2-8GHz. To the best of our knowledge, such high linearity is first measured at the frequency up to 8GHz. In addition, a 1×2 multi-mode interference 3dB-splitter, a photobleached refractive index taper and a quasi-vertical taper are used to reduce the optical insertion loss of the device.

**Index term:** Linear modulator, traveling wave electrode, electro-optic polymer, directional coupler, $\Delta\beta$ reversal, spurious free dynamic range

1. INTRODUCTION

Optical modulators in analog optical links are required to have high modulation efficiency, good linearity and large bandwidth. Existing commercial LiNibO$_3$ Mach-Zehnder (MZ) modulators have intrinsic drawbacks in linearity to provide high fidelity communication. When multiple tones of signals ($f_1$ and $f_2$) are simultaneously carried over a link, nonlinear intermodulation distortion signals are generated. The third-order intermodulation distortions (IMD3s), which are the interactions between fundamental frequencies and harmonics and occur at ($2f_1$- $f_2$) and ($2f_2$- $f_1$), are considered the most troublesome among all the nonlinear distortions because they usually fall at frequencies very close to fundamental frequencies and are in the pass band of the system. The spurious free dynamic range (SFDR) is defined as the dynamic range between the smallest signal that can be detected in a system and the largest signal that can be introduced into the system without creating detectable distortions in the bandwidth of concern [1]. The SFDR of high frequency analog optical links is limited by the system noise and the nonlinearity of modulation process. In order to improve the linearity of modulators, various efforts had been taken either electronically [2, 3] or optically [4-6] to suppress the IMD3s. A shortcoming common in all these linearization techniques is that the improved linearity is achieved at the expense of sacrificing the simple device design. A bias-free Y-fed directional coupler (YFDC) modulator, on the other hand, had been proven to possess a highly linear transfer function without loss of the simplicity of device design [7]. Even higher linearity is achievable when YFDC is incorporated with $\Delta\beta$-reversal technique [8, 9]. We previously demonstrated a polymer based 2-domain YFDC modulator with the $\Delta\beta$-reversal at low modulation frequencies as a proof of concept [10], where we achieved the SFDR of 119dB/Hz$^{2/3}$ with 11dB enhancement over the conventional MZ modulator.

In addition to the requirement of high linearity, the bandwidth is another important factor in evaluating the performance of a modulator. The first demonstration of optical modulation at GHz frequencies was done on a traveling wave electro-optic (EO) LiNibO$_3$ modulator in 1970s [11, 12]. A traveling wave modulator based on an EO polymer operating in the GHz frequency regime was demonstrated in 1992 [13]. Later on, a polymeric modulator operating over 100GHz was verified by groups at UCLA and USC [14]. Up until today the highest frequency that the polymer modulator can work at was demonstrated to be as high as 200GHz by Bell Laboratories [15]. Polymer EO modulators can offer several advantages over the mature LiNibO$_3$ modulators due to the special properties of polymer materials as below [16-18]. Excellent velocity matching between microwaves and optical waves can be achieved due to a close match between the refractive index of polymers at microwave and optical frequencies, enabling ultra-broad bandwidth operation. The intrinsic relatively low dielectric constant of polymers (2.5-4) also enables 50-ohm driving electrodes to be easily achieved. Polymers also have very large EO coefficients, $\gamma_{33}$, which is advantageous for sub-volt half-wave switching voltage ($V_\pi$) [19-21]. For example, CDL1/PMMA, an EO polymer with $\gamma_{33}$=60pm/V, was used to achieve $V_\pi$=0.8V [19]. Another EO polymer with a very large $\gamma_{33}$=306pm/V was developed through controlled molecular self-assembly and lattice hardening [22]. In comparison, the EO coefficient of LiNibO$_3$ is only about 30pm/V. In addition, the refractive index of polymers (1.6-1.7) is nearly matched to that of glass optical fibers (1.5-1.6), enabling small Fresnel reflection loss at interfaces in butt-coupling. Polymers can be highly transparent, and the absorption loss can be below 0.1dB/cm at all key communication wavelengths [17]. And also, polymers are spin-on films so they can be easily spincoated onto any substrate. Recently, EO modulations on extremely small geometrical footprints have been demonstrated by infiltrating EO polymer into slot waveguides [23] or slotted photonic crystal waveguides [24]. So far, the largest effective in-device $\gamma_{33}$=735pm/V and the smallest $V_\pi L$=0.44V•mm have been demonstrated by our group using EO polymer infiltrated silicon slotted photonic crystal waveguides [25]. Furthermore, compared to the difficult implementation of the domain inversion technique on LiNibO$_3$ [26], Δβ-reversal can be easily achieved by domain-inversion poling on EO polymers. Based on the above advantages of polymer materials, EO polymer modulators have shown great potentials for a variety of applications, such as telecommunication, analog-to-digital conversion, phased-array radar, and electrical-to-optical signal transduction. Now polymer based modulators with high reliability have been commercially available [27].

In this paper, we demonstrate an EO polymer based traveling wave directional coupler modulator with $\Delta\beta$-reversal to extend the high linearity performance to the GHz frequency regime. A traveling wave electrode with a unique design for RF microprobe coupling is fabricated with low microwave loss, characteristic impedance matching with 50Ω, and velocity matching between microwaves and optical waves, as well as smooth electric field profile transformation. The bandwidth-length product of 302GHz cm and the 3-dB bandwidth of 10GHz are achieved. The SFDR of 110±3dB/Hz$^{2/3}$ is measured over the modulation frequency of 2-8GHz. In addition, a 1×2 multi-mode interference (MMI) 3dB-splitter, a photobeached refractive index taper and a quasi-vertical taper, as well as a smooth silver ground electrode, are used to reduce the optical insertion loss of the device.

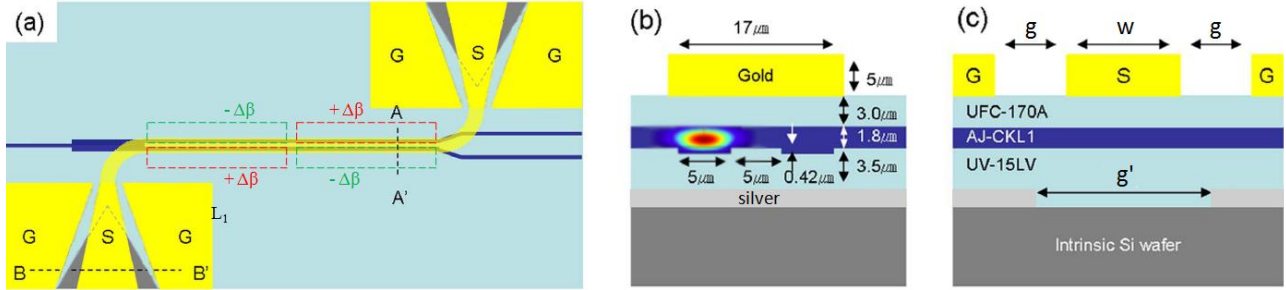

FIG. 1. (a) The schematic top view of the traveling wave MMI-fed directional coupler modulator with 2-domain-inversion. The red and green dashed lines indicate the area of EO polymer poled in opposite directions. (b) Cross section corresponding to A-A′ in (a), overlaid the optical mode profile in one arm. (c) Cross section corresponding to B-B′ in (a). (S: signal electrode, G: ground electrode).

## 2. DESIGN

### 2.1 Optical waveguide design

Fig. 1 (a) shows the schematic top view of our traveling wave MMI-fed directional coupler modulator. The cross section of the optical waveguides consisting of three layers of fluorinated polymers (bottom cladding: UV-15LV, core: AJ-CKL1/APC with $\gamma_{33}$=80pm/V, and top cladding: UFC-170A) is shown in Fig. 1 (b) and (c). Unlike the sine-squared transfer curve of the conventional MZ structure, a proper design of coupling length of directional coupler can provide a linear transfer function [2-6]. The linearity of the directional coupler can be further improved by applying $\Delta\beta$-reversal technique to suppress IMD3s [7-10]. Multiple-domain-inversion, which helps increase the linearity of directional coupler modulator, has been demonstrated by our group [28]. Considering the fabrication and poling complexity, in this paper we use a 2-domain-inversion directional coupler for demonstration. The directional coupler is divided into two domains, where EO polymer in the first domain is poled in the opposite direction with respect to that in the second domain. The push-pull configuration is also applied, in which the two arms of the directional coupler in each domain are poled in opposite directions, to double the EO effect. Finally, a single uniform modulation electric field applied by a traveling wave electrode can create $\Delta\beta$-reversal which is indicated by the dashed lines in Fig. 1 (a).

The IMD3 suppression of a directional coupler modulator is a sensitive function of the normalized interaction length ($S_i$), defined as the ratio of the interaction length ($L_i$) of i$^{th}$ section to the coupling length ($l_c$). Relative IMD3 suppression of a 2-domain directional coupler modulator can be graphically represented by plotting the calculated IMD3 suppressions on ($S_1$, $S_2$) plane [29]. $S_1 = S_2$ =2.86 provides excellent linearity as well as very high modulation depth [28] and is chosen for demonstration in this paper. For a directional coupler with the total interaction length ($L_1 + L_2$) of 2cm, its coupling length for TM mode should be 3496μm. This coupling length is matched by tuning the parameters of the trench waveguide, such as the core thickness and trench depth, using numerical methods [30]. The thickness of cladding is chosen to be 3.5μm and 3μm at the bottom and top, respectively, considering both the requirement of low driving and poling voltage and the prevention of optical absorption by metallic electrodes. The final cross section dimensions are shown in Fig. 1 (b). Based on fabrication experiences and actual measurements, the actual bottom width of a 420nm-deep trench fabricated by reactive ion etching (RIE) is about 4μm while the top width is still 5μm as designed. However, the calculation results show that the resulting coupling length deviation is only 0.55%, which can be explained by the fact that most of the optical power is distributed in the core layer, as shown in Fig. 1 (b), and that the field profile interaction happens at the top side of the two trenches which is un-affected.

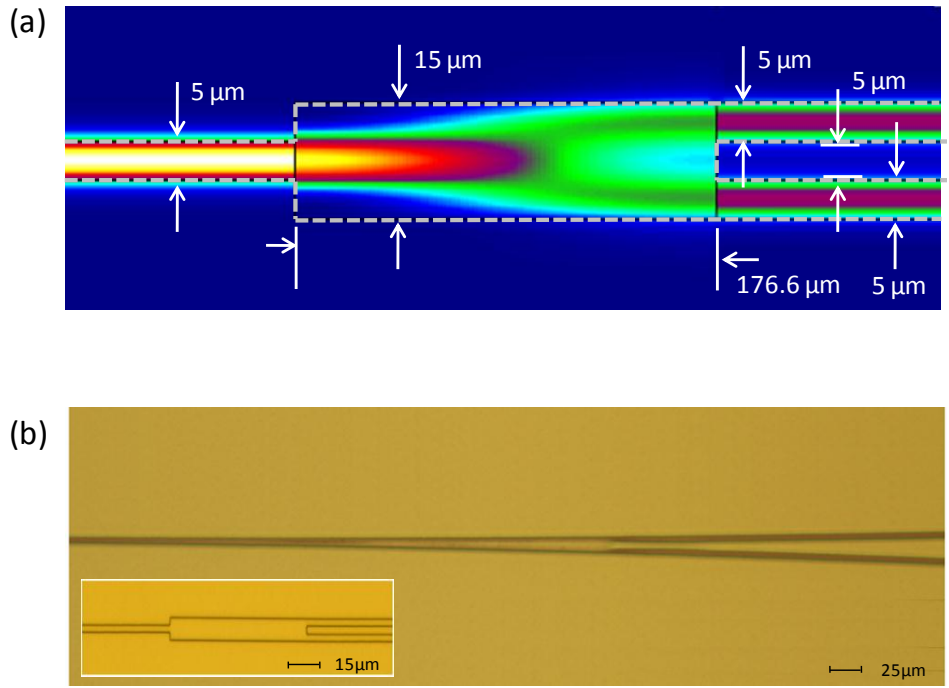

FIG. 2. (a) The top view of a 1×2 MMI 3-dB coupler, and the optical power distribution in this MMI coupler. (b) A blunt tip of a fabricated Y-junction due to fabrication limitations, compared to a fabricated MMI coupler shown in the inset.

A 1×2 MMI 3-dB coupler is designed to equally split the input optical power among two waveguides of a directional coupler as shown in Fig. 2 (a). The symmetric waveguide structure of the MMI-fed directional coupler is intrinsically bias-free; and the modulation is automatically set at 3-dB operation point regardless of the ambient temperature. The dimensions of MMI coupler and the optical power distribution in it are shown in Fig. 2 (a) and (b). The total power transmission of this MMI coupler is numerically calculated using eigenmode expansion method [31] to be as high as 94%. This MMI coupler has a large fabrication tolerance and is insensitive to the photolithography resolution. In comparison, as for previously used Y-junction [7], in practice the fabrication limitations in photolithography and etch resolutions usually lead to a blunt tip under a certain distance between the two waveguides (Fig. 2 (b)) and violates the adiabatic requirement, resulting in extra optical loss [32]. In addition, compared with the previous 1000 μm-long Y-junction [10], the MMI coupler is only 176.6 μm-long and is beneficial to decrease the device length. The inset of Fig. 2 (b) shows a microscope image of a fabricated MMI coupler compared with a traditionally used Y-junction.

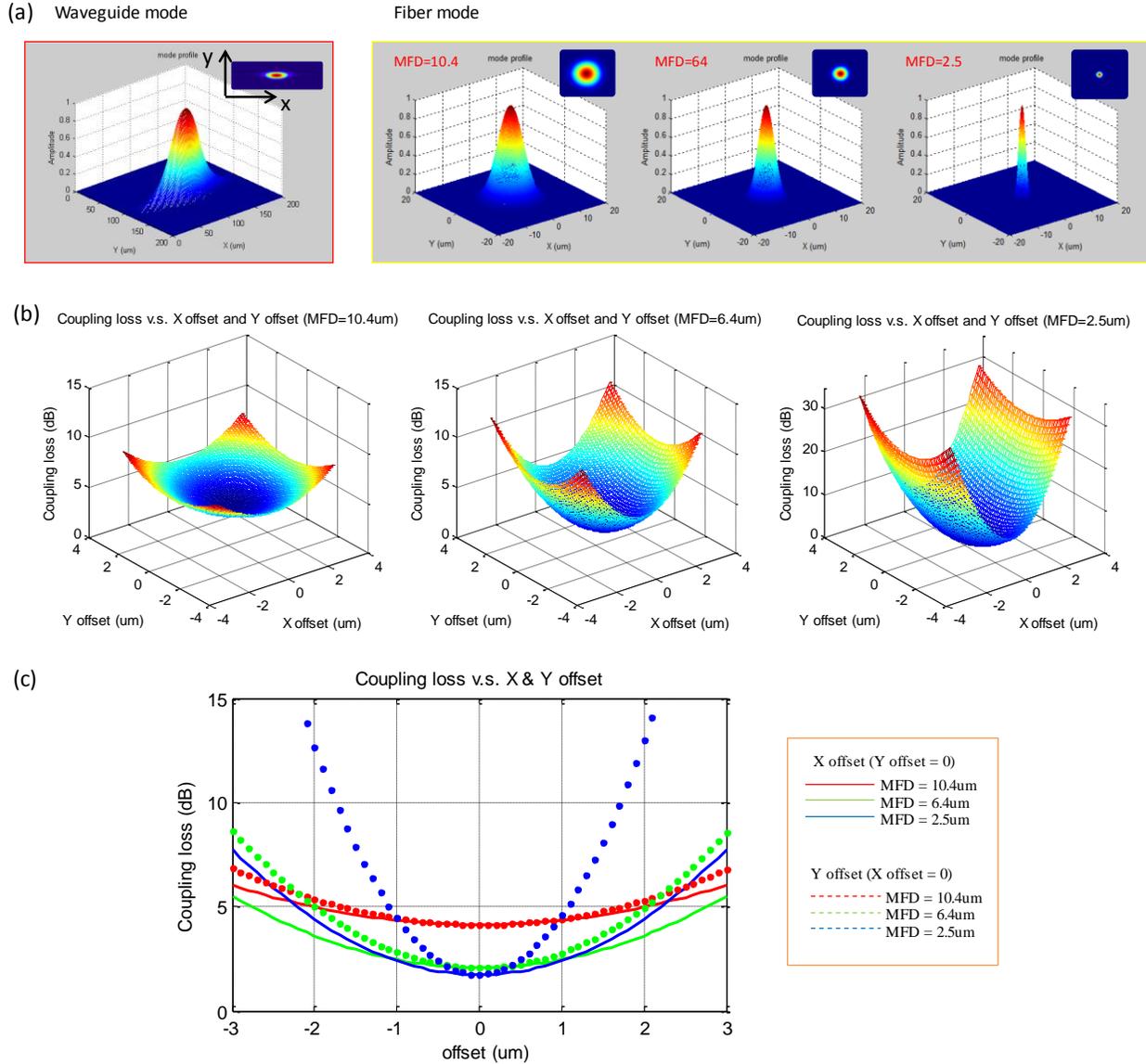

FIG. 3. (a) Optical mode profile in a polymer waveguide, compared with the optical mode profiles in I/O fibers with MFD of 10.4 μm, 6.4 μm and 2.5 μm. (b) The 3D perspective of the calculated coupling loss versus the misalignment in x and y direction, for using three different I/O fibers. (c) The 2D plot of the calculated coupling loss versus the misalignment in x and y direction. Red curves, green curves and blue curves represent the coupling loss using a fiber with MFD of 10.4 μm, 6.4 μm and 2.5 μm, respectively, and solid curves and dashed curves represent the coupling loss versus the misalignment in x and y directions, respectively.

A polymer trench waveguide is designed to support only a single mode. As shown in Fig. 3 (a), the TM mode profile of the designed polymer waveguide is calculated using finite element method [33], with the corresponding optical effective index of 1.599 at 1550nm. Fig. 3 (a) also shows the mode profiles of three single mode fibers with different mode field diameters (MFD). It can be seen that the mode profile of the polymer waveguide is in an elliptical geometry, with its major axis of about 7μm and minor axis of about 2.5μm, but that the mode profile of the normally used single mode fibers (e.g. SM980-5.8-125, Thorlabs) is in a circular geometry with MFD of 10.4μm. This mode size mismatch can lead to large optical loss in butt-couplings. To reduce such coupling loss, single mode fibers with MFD of 6.4μm (e.g. SM1500G80, Thorlabs) or lens fibers with MFD of 2.5μm (e.g. TSMJ-3U-1550-9/125-0.25-7-2.5-14-2, OZ Optics) can be used as replacements at the input/output (I/O) sides. The power-coupling loss using these three fibers are calculated by considering the overlap integral of mode profiles as well as Fresnel reflection loss at interfaces [34, 35]. Fig. 3 (b) shows the three-dimensional perspective of the coupling loss as a function of the spatial misalignment in x and y directions for these three fibers. To see a clear comparison of these three fibers, a two-dimensional plot of the same data is shown in Fig. 3(c). It can be seen that the lowest coupling loss, 1.7dB/facet, can be achieved using lens fibers but that there are very large variations versus misalignment along both x and y directions. It can be noticed that, using lens fibers, the coupling efficiency is more sensitive to the misalignment in y direction than that in x direction. Using single mode fibers with MFD of 6.4μm, a larger alignment tolerance can be achieved, while the peak coupling loss is 2.0dB/facet which is just a little higher than that of the lens fibers. Normally used single mode fibers with MFD of 10.5μm can provide the lowest misalignment sensitivity but the highest coupling loss which is up to 4.1dB/fact. Therefore, considering both coupling loss and misalignment tolerance, a single mode fiber with MFD of 6.4μm is finally chosen for our experiment.

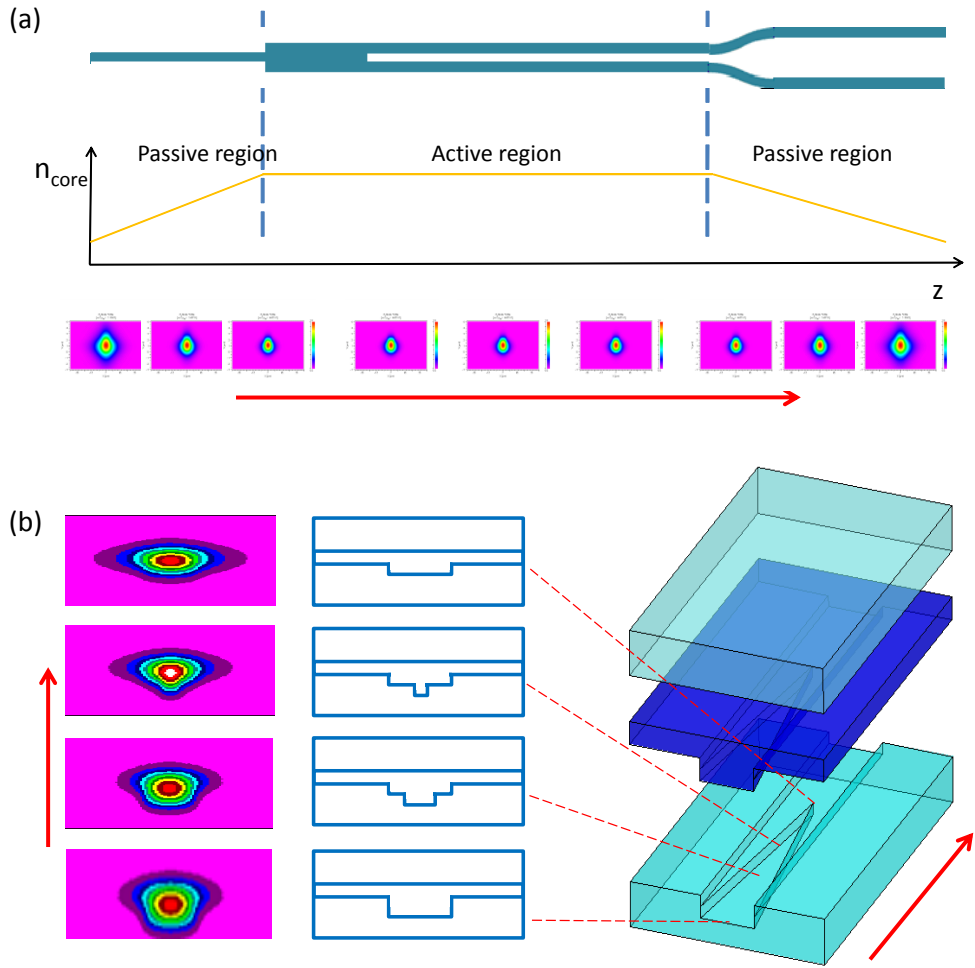

FIG. 4. (a) Refractive index tapers at the passive regions of the MMI-fed directional coupler. The index variation of the photobleached EO polymer in the core layer leads to the gradual change of optical mode size along the taper. (b) A quasi-vertical taper at one facet of polymer waveguide used for mode profile transformation in vertical direction. The red arrows indicate the beam propagation direction.

Another way to reduce the optical coupling loss due to mode size mismatch is to design a taper structure. Here, refractive index tapers are designed at the passive regions of the waveguides so that the optical mode profile at the I/O ends of the polymer waveguide can better match that of the I/O optical fibers. The working mechanism is shown in Fig. 4 (a). Refractive index variation of EO polymer core at the passive regions of the waveguide is created by UV photobleaching method, using a gray-scale photomask or discrete step mask-shifting scheme [36]. This refractive index variation leads to the gradual change of optical mode size along the taper, so that the waveguide modes at the facets are large enough to match that of I/O fibers. In addition, to minimize the severe mode size mismatch in vertical direction that can be seen in Fig. 3 (a), a quasi-vertical taper structure [37] is designed as shown in Fig. 4 (b). It can be fabricated by standard photolithography and RIE twice. After a trench is etched on the bottom cladding polymer, a V-shape groove is etched again into the trench near the I/O facets. This structure works as an optical mode transformer. Because the trenches are deeper at the facets than in the active regions of waveguide, the waveguide mode size in vertical direction becomes larger at the facets and can better match the I/O fiber mode. Numerical calculations using beam propagation method [38] show that the combination of these two tapers can significantly reduce the optical coupling loss by 3dB/facet.

Other than the coupling loss, the roughness of polymer waveguide sidewalls usually causes large scattering loss when light propagates in the waveguide. Thus, silver is selected as the ground electrode material and its smooth surface helps reduce waveguide sidewall roughness originating from the scattering of UV light in photolithography. Compared to other metals, silver is also beneficial to suppress the microwave conductor loss owing to its very low resistivity.

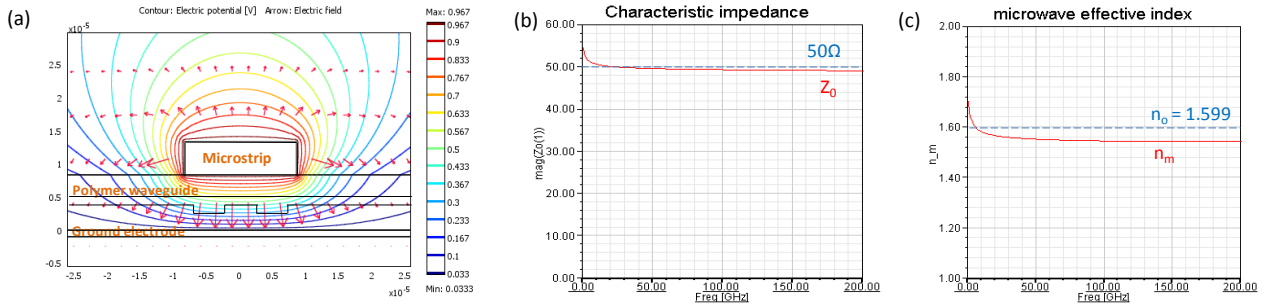

FIG. 5. (a) The schematic cross section of a microstrip line with design parameters overlaid the contour of the normalized electric potential. The red arrows indicate the direction of electric field. (b) The characteristic impedance of the microstrip line over the frequency range 1-200GHz. The solid red curve indicates the characteristic impedance and the dashed blue line indicates the 50Ω. (c) The microwave effective index of the microstrip line over the frequency range 1-200GHz. The solid red curve indicates the microwave effective index and the dashed blue line indicates the optical effective index of 1.599.

### 2.2 Traveling wave electrode design

To extend the highly linear modulation to GHz frequency regime, a traveling wave electrode is necessary. Some basic requirements for the design of a high-speed traveling wave electrode are [39]: (i) impedance matching between the microwave guides and external electrical connectors, (ii) velocity matching between the microwave and optical signals, and (iii) low electrical loss in the microwave guides. In addition to the above requirements, when designing a transition between different types of microwave guides, electric field matching [40, 41] should also be a concern in order to reduce the microwave coupling loss. In our device structure, polymer is considering the alignment of modulation field with the direction of the $\gamma_{33}$ in the poled EO polymer film, which is in vertical direction in our device configuration; therefore, a microstrip line is a natural choice for the best alignment. Fig. 5 (a) shows the schematic cross section of the designed gold microstrip line overlaid the contour of the normalized electric potential calculated by finite element method [33]. It can be seen that both arms of the directional coupler waveguide are under the effect of a uniform modulation field between the microstrip line and the ground electrode and hence the overlap integral between the optical mode and the RF modulation field can be maximized. In quasi-static analysis, the characteristic impedance $Z_0$ and the microwave effective index $n_m$ of a transmission line can be expressed as [39, 42]

$$Z_0 = \frac{1}{c \sqrt{(C \, C_a)}} \tag{1}$$

$$n_m = \left(\frac{C}{C_a}\right)^{1/2} \tag{2}$$

where $C_a$ is the capacitance per unit length of the electrode structure with the dielectrics replaced by air, $C$ is the capacitance per unit length with the dielectrics present, and $c$ is the speed of light in vacuum. The frequency-dependent characteristic impedance and microwave effective index can be numerically calculated using finite element method [43] to match 50Ω and optical effective index of 1.599, respectively. Conductor loss and dielectric loss are considered in the calculation so that the results are accurate enough and close to the real case. Given the relative dielectric constant $\varepsilon_r = 3.2$, the gap between top and bottom electrodes $h = 8.3\,\mu m$ and the microstrip thickness $t = 5\,\mu m$ from the waveguide dimensions and the fabrication conditions, the characteristic impedance of 50Ω can be matched when the microstrip width w=17 μm. As shown in Fig. 5 (b) and (c), over the frequency range 1-200GHz, the characteristic impedance varies within 49-54.5Ω and the microwave effective index varies within 1.54-1.7. It can be noticed from Fig. 5 (b) that the characteristic impedance at low frequencies is relatively higher than that at high frequencies. This is because the internal inductance of the microstrip line decreases with frequency and becomes negligible when skin effect kicks in. Based on fabrication experience and actual measurements, the electroplated gold microstirp line does not have a perfect vertical sidewall but a wall angle of 84°; however, the variation of $Z_0$ and $n_m$ due to this wall angle are calculated to be within 1Ω and 0.005, respectively, which can be negligible. The bandwidth-length product due to the velocity mismatch can be calculated as [11, 39, 44, 45]

$$f \cdot L \cong \frac{1.9c}{\pi |n_m - n_o|} \tag{3}$$

where $f$ is the modulation frequency, $L$ is the interaction length, $c$ is the speed of light in vacuum, $n_m$ is the microwave effective index of the microstrip line, and $n_o$ is the optical effective refractive index of polymer waveguide. Using Equation 3, the bandwidth-length product can be theoretically calculated to be up to 306GHz cm, corresponding to a modulation frequency limit of 153GHz for a 2cm-long microstrip line.

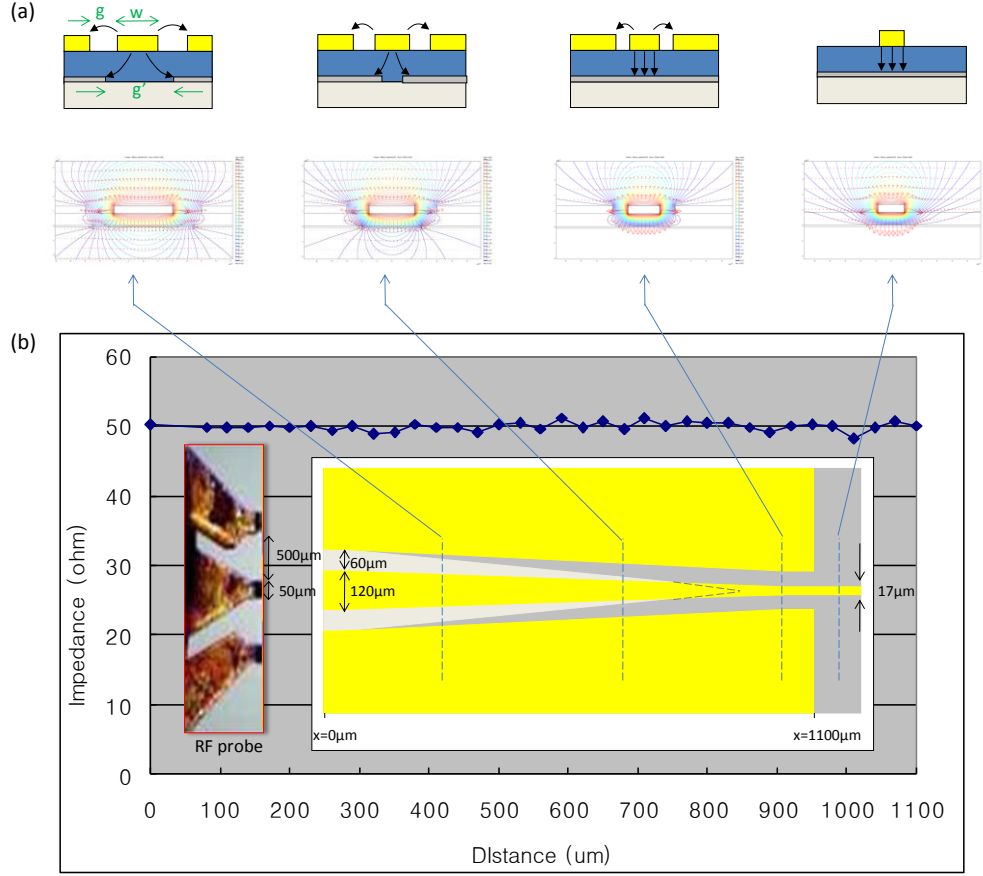

FIG. 6. (a) The schematic of a smooth transformation of electric field profile in the CPW-to-microstrip transition at the input end of traveling wave electrode. The corresponding distribution of electric field and electric potential along this transition taper is calculated with finite element method. (b) The top view of the quasi-CPW taper, matching the size of a microprobe. The characteristic impedance (at 10GHz) is matched with 50Ω along the transition direction.

To couple the RF power from a GSG microprobe (e.g. ACP40-GSG-250, Cascade Microtech, probe tip width: 50μm, pitch: 500μm, as shown in Fig. 6 (b)) into the 17μm-wide microstrip line with minimum coupling loss, a 1.1mm-long quasi-coplanar waveguide (CPW) taper is designed at input end, as shown in Fig. 6 (b) . The top width and gap of the coplanar waveguide (w and g in Fig. 6 (a)) are gradually changed along the taper to match the dimension of a RF microprobe. In the transition between the CPW and the microstrip line, the electric field profiles of these two microwave guides should be matched to reduce microwave coupling loss. Therefore, unlike the conventional CPW, the ground electrode under the taper is partially removed and the bottom gap (g´ in Fig. 6 (a)) is gradually tuned along the taper based on ground shaping technique [40, 41], so that there is a smooth transformation of electric field profile in the CPW-to-microstrip transition [33] while the 50Ω is matched at all points along the transition direction [43], as shown in Fig. 6. At the output end, a similar microstrip-to-CPW transition taper is designed to couple the RF power from the microstrip line to another GSG microprobe (Fig. 1 (a)). What is more, for the design to be valid, the resistivity of silicon substrate should be sufficiently high (typical 1 kΩ cm or higher). Otherwise, the finite conductivity of silicon substrate allows the formation of microstrip mode between the signal electrode and silicon substrate, and this microstrip mode would become dominant at wide part of the taper and hinder 50Ω matching. Therefore, an intrinsic silicon wafer with ultra-high resistivity (6-10kΩ•cm) should be used as the substrate of our device.

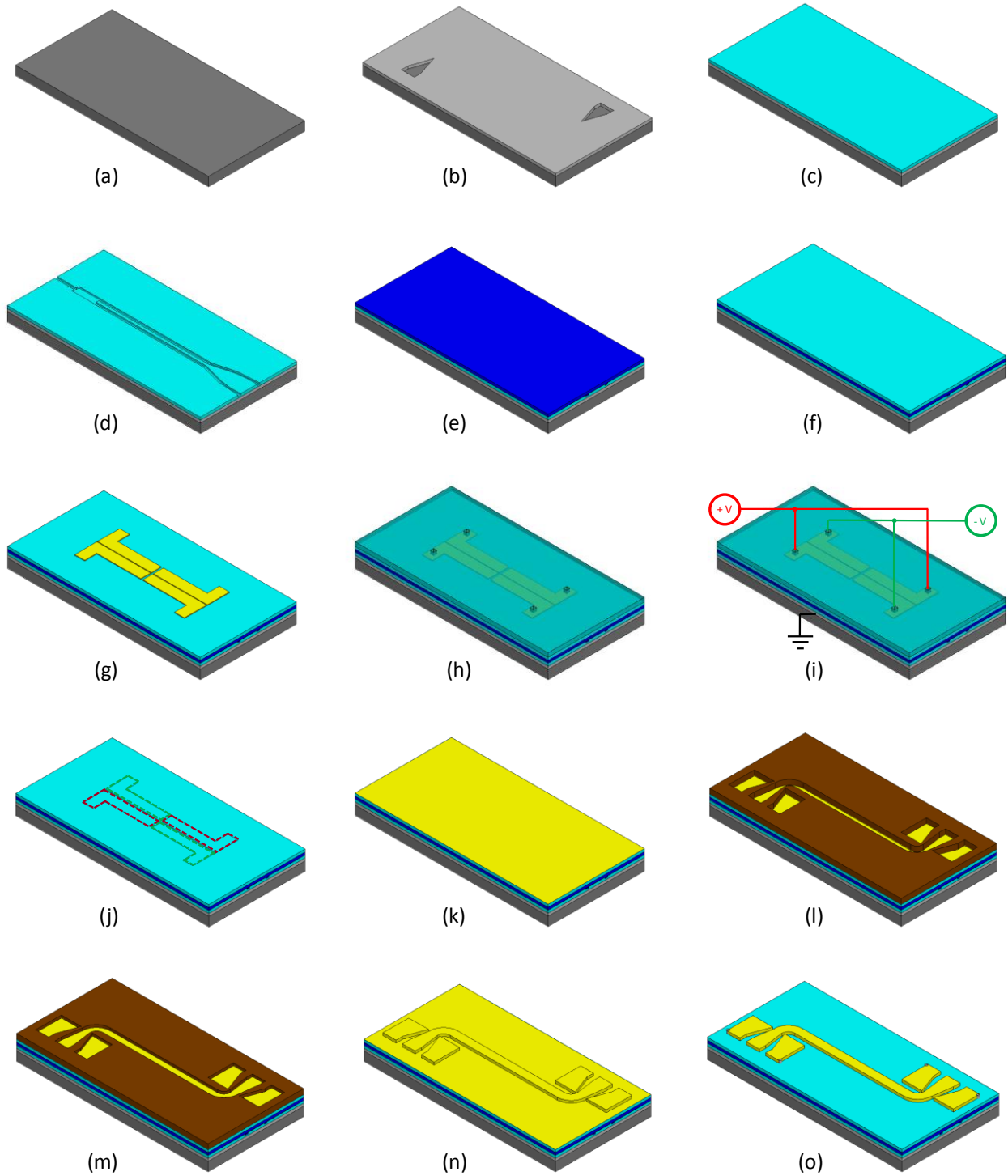

FIG. 7. Fabrication process flow. (a) An ultra-high resistivity silicon wafer. (b) Ground electrode deposition and patterning. (c) Bottom cladding deposition. (d) Waveguide patterning. (e) EO polymer formulation and deposition. (f) Top cladding deposition. (g) Poling electrode deposition and patterning. (h) Protection layer deposition and patterning. (i) Poling. (j) Removal of protection layer and poling electrode. (k) Seed layer deposition. (l) Buffer mask deposition and patterning. (m) Traveling wave electrode electroplating. (n) Buffer mask removal. (o) Seed layer removal and vias drilling.

## 3. FABRICATION

Fig. 7 illustrates the fabrication process flow. The device is fabricated on an ultra-high resistivity silicon wafer. A 1 μm-thick silver film is deposited by electron-beam evaporation and then patterned using lift-off process, to serve as the ground electrode for poling process as well as for RF transmission. A polymer trench waveguide is fabricated by spincoating, photolithography and RIE, in which the EO polymer is formulated by doping 25wt% of AJ-CKL1chromophore into amorphous polycarbonate (APC). 150nm-thick gold poling electrodes are deposited by electron-beam evaporation and patterned by lift-off process. 300nm-thick silicon dioxide is deposited by electron-beam evaporation to cover the entire surface of the device as a protection layer. Then contact windows are opened on the silicon dioxide using photolithography and wet etching method, so that the electrodes can be exposed to the probe needles in the following poling process.

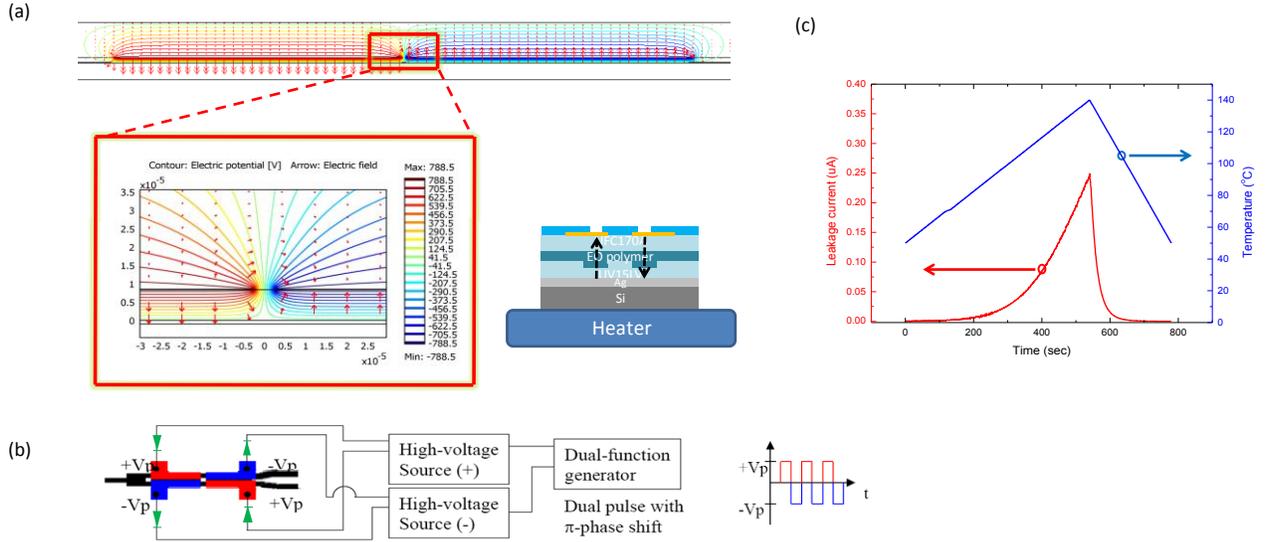

FIG. 8. (a) The cross section of poling electrodes above the polymer waveguide overlaid the electric potential distribution in push-pull poling configuration. (b) The schematic of push-pull, 2-domain-inversion, alternating-pulse poling. (c) The temperature dependence of leakage current during the poling time.

Poling is the most important step throughout the entire process since the EO coefficient of a device is determined by poling efficiency [46, 47]. Among several developed poling techniques [48-52], we employ thermally-assisted electric-field contact-poling [53] in this work. In the push-pull configuration, two adjacent poling electrodes above the two arms of directional coupler have opposite polarities and thus the electric field formed between electrodes is very strong, as shown in Fig. 8 (a). Given the electrode separation of 5 μm, the polymer waveguide thickness of 8.3 μm and the poling electric field of typically ±100V/μm applied vertically across the polymer waveguide, the maximum electric field between two adjacent electrodes is calculated to be over 300V/μm [33]. This increases the probability of dielectric breakdown which can easily damage the device. To prevent this, the deposited thick silicon dioxide serves as a protection layer, as shown in Fig. 8 (a), due to its good insolating property and high dielectric strength (up to 1000V/μm). Experimental tests show that the poling electric field up to 150V/μm can be applied on our device structure at the glass transition temperature ($T_g$=140°C) of EO polymer without dielectric breakdown. This can significantly increase the poling efficiency. In addition, alternating-pulse poling technique [54, 55] is used to further prevent dielectric breakdown. As shown in Fig. 8 (b), the positive and negative voltage sources are controlled by dual pulse with π-phase shift from a dual-function generator, so that two opposite polarities are not applied at the same time. Four diodes are used as clampers. Based on testing experience, the frequency of the alternating pulses should be set to be 1-10Hz to avoid dielectric breakdown. During the poling process, the temperature is controlled to increase from room temperature to $T_g$ and then quickly decrease back to room temperature. Throughout the entire poling process, leakage current is monitored by a picoammeter. A current-limiting resistor and two back-to-back diodes are used in the circuit connection to protect the picoammeter from being damaged by any unexpected breakdown-induced high current. Fig. 8 (c) shows a leakage current curve depending on the controlled temperature during the poling time.

After poling is done, the silicon dioxide layer and the poling electrodes are removed by wet etching method. A 50nm-thick gold seed layer with 5nm-thick chromium adhesion buffer is then deposited above the polymer waveguide by electron-beam evaporation. The buffer mask for the traveling wave electrode is patterned on 10μm-thick AZ-9260 photoresist by photolithography. A 5μm-thick gold film is electroplated using Techni-Gold 25ES electrolyte. A constant current of 8mA is used in the entire gold electroplating process. For the electroplating area of about 11cm$^2$, the corresponding current density is as low as 0.73mA/cm$^2$ which enables a uniform gold thickness. The conductivity of the electroplated gold film is measured to be $2.2 \times 10^7$S/m. The coplanar and ground electrodes are then connected with silver epoxy through via-holes. Finally, the device is diced and the waveguide facets are polished.

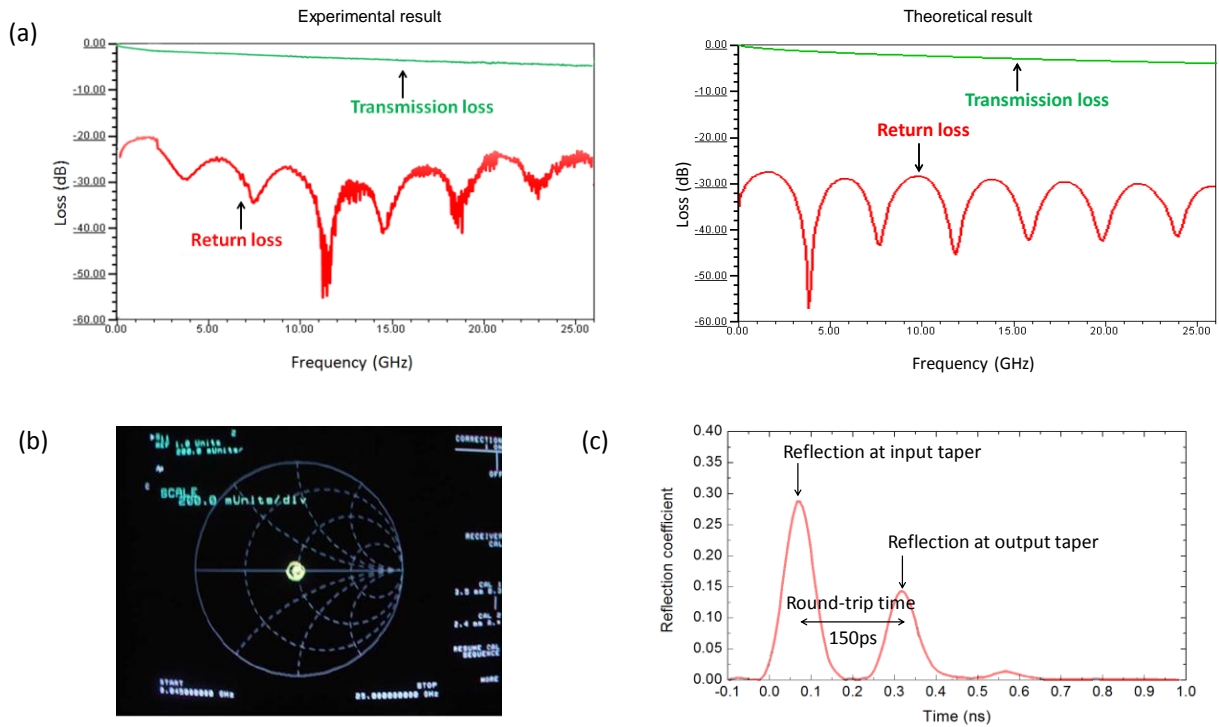

FIG. 9. (a) The measured transmission loss and return loss of the fabricated traveling wave electrode over the frequency range 1-26GHz (left side), almost matching the theoretical calculations (right side). (b) The measured characteristic impedance of the fabricated traveling wave electrode is well centered at 50Ω on Smith Chart, indicating impedance matching. (c) The time domain measurement of the reflection loss, for the demonstration of velocity matching.

## 4. TESTING

### 4.1 Electrode characterization

The performance of the fabricated traveling wave electrode is characterized by a vector network analyzer (HP 8510C). Two air coplanar probes (ACP40-GSG-250, Cascade Microtech) are used to couple RF power into and out of the tapered quasi-coplanar waveguides. The measured microwave loss of the traveling wave electrode over the frequency range 1-26GHz (upper frequency limited by equipment) is presented on the left side in Fig. 9 (a). For reference, the theoretically calculated electrode loss using finite element method [43] is shown on the right side in Fig. 9 (a). It can be seen that the measured transmission loss is proportional to the square root of frequency, implying that the microwave loss is dominated by the conductor loss (skin effect loss) of the electrode [56, 57] which is measured to be 0.65±0.05dB/cm/GHz$^{1/2}$. The 3-dB electrical bandwidth measured from transmission loss curve is 10GHz, nearly the same value as that from the theoretical calculation in which the actual conductor loss of the electroplated gold electrode has been considered. This bandwidth is limited by the relatively low conductivity of the poorly electroplated gold electrode and can be enhanced by improving the electroplating quality. The measured return loss is well below -20dB. This low return loss is mainly due to the excellent impedance matching as well as the smooth electric field transformation in the CPW-microstrip-CPW transition. It can be noticed that this value is still higher than the theoretical result (<-27dB), probably due to the fabrication imperfection. The periodic ripples in the return loss curve are attributed to the RF Fabry-Perot effect. It is shown in Fig. 9 (b) that the characteristic impedance is well centered at 50Ω on Smith chart, indicating impedance matching. The velocity matching between microwaves and optical waves is evaluated by the time domain measurement of the return loss, as shown in Fig. 9 (c). The effective relative dielectric constant of the microstrip line is measured to be 2.76 and the resulting index mismatch between microwave sand optical waves is 0.06. Then the bandwidth-length product due to this velocity mismatch can be calculated by Equation (3) to be 302GHz cm, so the modulation frequency limit corresponding to 2cm interaction length would be 151GHz, which matches the theoretical calculation result (153GHz from Fig. 5 (c)) pretty well.

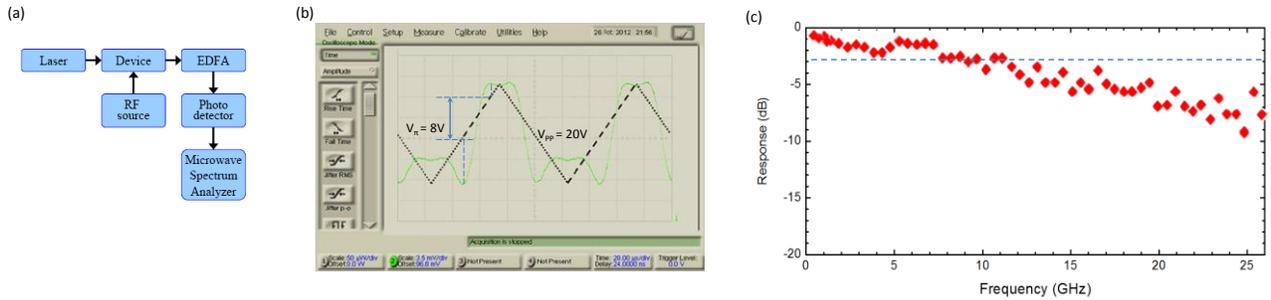

FIG. 10. (a) The schematic of testing system for small signal optical modulation measurement. (b) Transfer function of over-modulation with $V_{pp}$=40V at 10kHz (wavelength=1550nm). The switching voltage is measured to be 16.5V. (c) The frequency response of the small signal modulation measured at 4% modulation depth. The 3-dB bandwidth is measured to be 10GHz.

**4.2 Small signal optical modulation measurement**

The frequency response of the device is evaluated by the small signal optical modulation measured at 4% modulation depth. The testing system is shown in Fig. 10 (a). TM-polarized light with 1550nm wavelength from a tunable laser (Santec ML-200, Santec Corp.) is butt-coupled into the waveguide through a single mode fiber. The measured optical insertion loss is 16 dB, which includes propagation loss of 9dB (absorption loss of 2dB/cm for AJ-CKL1 times the total device length of 3cm, scattering loss of 1dB/cm times 3cm), coupling loss of 6dB (3dB/fact times 2 facets), and 1 dB loss from the MMI splitter. This relatively high loss is attributed to the roughness of the 3-cm long waveguide sidewalls generated in RIE process and the roughness of input and output waveguide facets. To measure the switching voltage ($V_\pi$), a testing RF signal with $V_{pp}$=20V at 10kHz is used. The transfer function of an over-modulation test is shown in Fig. 10 (b). The switching voltage is measured to be 8V at 10kHz, which is a little high probably due to the low poling efficiency of EO polymer and electrode loss. For the small signal optical modulation, RF signal from HP 83651B is fed into the traveling wave electrode through a GSG microprobe. The modulated optical signal is boosted by an erbium doped fiber amplifier (Intelligain, Bay Spec Inc.), converted to electrical signal by a photodiode (DSC-R409, Discovery Semiconductors Inc.), and then measured by a microwave spectrum analyzer (HP 8560E). The frequency response measured at 4% modulation depth is presented in Fig. 10 (c), from which the 3-dB bandwidth of the device can be found to be 10GHz. This bandwidth is mainly limited by conductor loss of the traveling wave electrode.

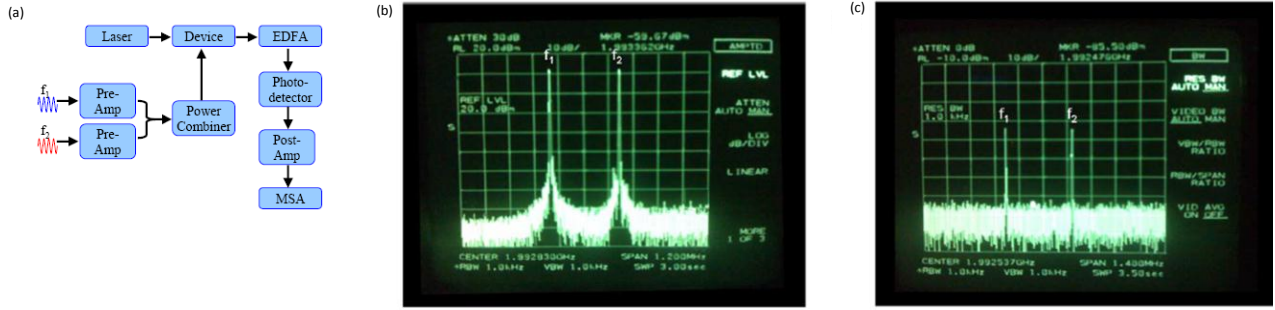

FIG. 11. (a) The schematic of system for two-tone test. (b) Input two-tone signals ($f_1$ and $f_2$) centered at 1.9928 GHz with 330 kHz tone-interval. (c) Measured output fundamental signals.

### 4.3 Linearity evaluation

A two-tone test is performed to evaluate the linearity of the device. The testing system is illustrated in Fig. 11 (a). HP 8620C sweep oscillator is used as the second RF source for the two-tone input signals. Agilent 83020A and HP8449B are used as pre- and post-RF amplifiers, respectively. The two input RF signals are combined by a coaxial two-way RF power combiner (RFLT2W1G04G, RF-Lambda). New Focus model-1014 is used for optical-to-electrical conversion of the modulated signal. The two-tone input signals and the resulting output signals are shown in Fig. 11 (b) and (c), respectively. IMD3 signals, which are supposed to appear at one tone-interval away from the fundamental signals if present, are not observed in Fig. 11 (c). A possible reason is that the IMD3 signals are well suppressed and buried under the noise floor at this modulation depth. The power level of the two-tone input signals is 12dBm as shown in Fig. 11 (b), which is the maximum level available in our two-tone test setup, and this power level translates into the modulation depth of 15%. The simulation result in [28] predicts the IMD3 suppression at 15% modulation depth to be 74dB and the corresponding experimental result in [10] is 69dB, which is a reasonable value considering the fabrication and measurement errors. Neglecting the performance degradation due to microwave loss and velocity mismatch, IMD3 signals would be 30dB below the noise floor in Fig. 11 (b) at 1kHz bandwidth resolution.

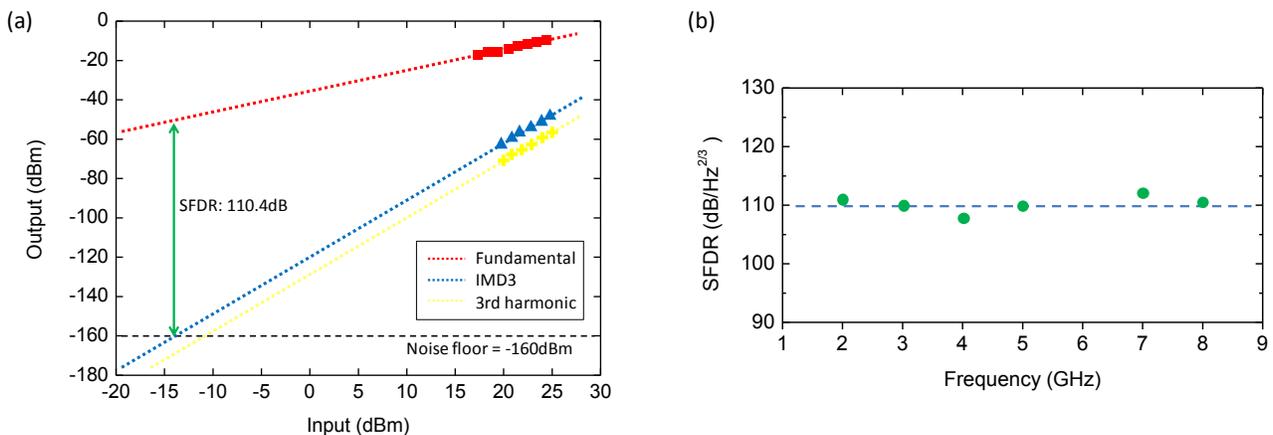

FIG. 12. (a) The plot of fundamental and third-order intermodulation distortion signals measured at 8GHz. (b) Spurious free dynamic range measured at 2-8GHz.

Since the IMD3 suppression of the fabricated device is out of the measurable range in our two-tone test setup, SFDR is evaluated through an indirect method. It is known that, with the same modulation depth for both tones, IMD3 is three times or 9.54dB higher than the third harmonic distortion [58]. A mono-tone test is done under the same conditions as the two-tone test. The power level of mono-tone input signal is extended up to 29dBm by combining the RF source (HP 83651B) with the pre-amplifier (Agilent 83020A). It is found that the third harmonic distortion of our device comes in the detectable range at the mono-tone input signal level above 20dBm. The IMD3 signals are obtained by adding 9.54dB to the measured third harmonic distortion signals. The SFDR is measured by extrapolating the IMD3 plot to find an intercept point with the noise floor and then measuring the difference with the extrapolated fundamental signal as illustrated in Fig. 12 (a). Considering the relative intensity noise (RIN) of the distributed feedback (DFB) laser and the shot noise of the photodiode, it is very difficult to achieve a noise floor below -145dBm in real analog optical links [28]. However, laboratory test results in most literatures are frequently presented assuming the noise floor at -160dBm considering the typical fiber-optic link parameters [2, 59, 60]. Using -160dBm as noise floor, our measured SFDR is within $110\pm3$dB/Hz$^{2/3}$ over the modulation frequency range 2-8GHz as shown in Fig. 12 (b). The low end frequency is determined by the operation range (2-26.5GHz) of the pre-amplifier (Agilent 83020A) and the high end is limited to 8GHz because the third harmonic of the modulation frequency above 8GHz goes beyond the scope (~26.5GHz) of the microwave spectrum analyzer. The SFDR at 6GHz is missing due to the irregular gain of the post-amplifier at 18GHz. As a comparison, Schaffner et al. reported the SFDR of 109.6dB/Hz$^{2/3}$ at 1 GHz with a LiNibO$_3$ directional coupler modulator which is linearized by adding passive bias sections [1]. In their measurement, however, the noise floor was set at -171dBm, which offers 7.3dB extra dynamic range compared with the noise floor at -160dBm. Hung et al. achieved even higher SFDR of 115.5dB/Hz$^{2/3}$ at 3GHz with a linearized polymeric directional coupler modulator by subtracting the distortions of the measurement system [60]. Here our SFDR of $110\pm3$dB/Hz$^{2/3}$ includes the distortions from the entire measurement system as well as the device. To the best of our knowledge, such high linearity is first measured at a frequency up to 8GHz.

## 5. CONCLUSION

We have demonstrated a linearized traveling wave MMI-fed directional coupler modulator based on EO polymer. Domain-inversion poling is applied to implement the Δβ-reversal technique. The traveling wave electrode is evaluated to be functional up to 151GHz for our device design due to the excellent velocity matching between microwaves and optical waves. The SFDR of $110\pm3$dB/Hz$^{2/3}$ is achieved over the modulation frequency of 2-8GHz. The measured 3-dB bandwidth of the device is 10GHz, which is mainly limited by conductor loss and needs further improvement for practical applications. The optical loss still needs to be further suppressed for application, and this will be improved in our future work by choosing low loss sol-gel passive materials and by using wet etching method for waveguide fabrication [61].

**ACKNOWLEDGEMENT**

Financial supports from the Defense Advanced Research Projects Agency (DARPA) under Contract No. SBIR W31P4Q-08-C-0160 monitored by Dr. Devnand Shenoy is acknowledged.

**REFERENCES**

[1] J. H. Schaffner, J. F. Lam, C. J. Gaeta, G. L. Tangonan, R. L. Joyce, M. L. Farwell, and W. S. C. Chang, "Spur-free dynamic range measurements of a fiber optic link with traveling wave linearized directional coupler modulators," *Photonics Technology Letters, IEEE,* vol. 6, pp. 273-275, 1994.

[2] R. B. Childs and V. A. O'Byrne, "Predistortion linearization of directly modulated DFB lasers and external modulators for AM video transmission," in *Proc. Tech. Dig. Opt. Fiber Commun. Conf.,*, San Francisco, CA, 1990.

[3] R. M. D. E. Ridder and S. Korotky, "Feedforward compensation of integrated optic modulator distortion," in *Proc. Tech. Dig. Opt. Fiber Commun. Conf.*, San Francisco, CA, 1990.

[4] L. M. Johnson and H. Roussell, "Reduction intermodulation distortion in interferometric optical modulators," *Optics letters,* vol. 13, pp. 928-930, 1988.

[5] S. K. Korotky and R. De Ridder, "Dual parallel modulation schemes for low-distortion analog optical transmission," *Selected Areas in Communications, IEEE Journal on,* vol. 8, pp. 1377-1381, 1990.

[6] M. L. Farwell, Z. Q. Lin, E. Wooten, and W. Chang, "An electrooptic intensity modulator with improved linearity," *Photonics Technology Letters, IEEE,* vol. 3, pp. 792-795, 1991.

[7] S. Thaniyavarn, "Modified 1×2 directional coupler waveguide modulator," *Electronics Letters,* vol. 22, pp. 941-942, 1986.

[8] H. Kogelnik and R. V. Schmidt, "Switched directional couplers with alternating Δβ," *Quantum Electronics, IEEE Journal of,* vol. 12, pp. 396-401, 1976.

[9] R. F. Tavlykaev and R. V. Ramaswamy, "Highly linear Y-fed directional coupler modulator with low intermodulation distortion," *Lightwave Technology, Journal of,* vol. 17, pp. 282-291, 1999.

[10] B. Lee, C. Y. Lin, A. X. Wang, R. Dinu, and R. T. Chen, "Linearized electro-optic modulators based on a two-section Y-fed directional coupler," *Applied optics,* vol. 49, pp. 6485-6488, 2010.

[11] M. Izutsu, Y. Yamane, and T. Sueta, "Broad-band traveling-wave modulator using a LiNbO3 optical waveguide," *Quantum Electronics, IEEE Journal of,* vol. 13, pp. 287-290, 1977.

[12] M. Izutsu, T. Itoh, and T. Sueta, "10 GHz bandwidth traveling-wave LiNbO3 optical waveguide modulator," *Quantum Electronics, IEEE Journal of,* vol. 14, pp. 394-395, 1978.

[13] C. Teng, "Traveling‐wave polymeric optical intensity modulator with more than 40 GHz of 3‐dB electrical bandwidth," *Applied Physics Letters,* vol. 60, pp. 1538-1540, 1992.

[14] D. Chen, H. R. Fetterman, A. Chen, W. H. Steier, L. R. Dalton, W. Wang, and Y. Shi, "Demonstration of 110 GHz electro-optic polymer modulators," *Applied Physics Letters,* vol. 70, p. 3335, 1997.

[15] M. Lee, H. E. Katz, C. Erben, D. M. Gill, P. Gopalan, J. D. Heber, and D. J. McGee, "Broadband modulation of light by using an electro-optic polymer," *Science,* vol. 298, pp. 1401-1403, 2002.

[16] L. R. Dalton, A. W. Harper, B. Wu, R. Ghosn, J. Laquindanum, Z. Liang, A. Hubbel, and C. Xu, "Polymeric Electro‐Optic Modulators: Matereials synthesis and processing," *Advanced Materials,* vol. 7, pp. 519-540, 1995.

[17] H. Ma, A. K. Y. Jen, and L. R. Dalton, "Polymer-based optical waveguides: materials, processing, and devices," *Advanced Materials,* vol. 14, pp. 1339-1365, 2002.

[18] M. C. Oh, H. Zhang, C. Zhang, H. Erlig, Y. Chang, B. Tsap, D. Chang, A. Szep, W. H. Steier, and H. R. Fetterman, "Recent advances in electrooptic polymer modulators incorporating highly nonlinear chromophore," *Selected Topics in Quantum Electronics, IEEE Journal of,* vol. 7, pp. 826-835, 2001.

[19] Y. Shi, C. Zhang, H. Zhang, J. H. Bechtel, L. R. Dalton, B. H. Robinson, and W. H. Steier, "Low (sub-1-volt) halfwave voltage polymeric electro-optic modulators achieved by controlling chromophore shape," *Science,* vol. 288, pp. 119-122, 2000.

[20] Y. Enami, C. Derose, D. Mathine, C. Loychik, C. Greenlee, R. Norwood, T. Kim, J. Luo, Y. Tian, and A. K. Y. Jen, "Hybrid polymer/sol–gel waveguide modulators with exceptionally large electro–optic coefficients," *Nature Photonics,* vol. 1, pp. 180-185, 2007.

[21] Y. Enami, D. Mathine, C. DeRose, R. Norwood, J. Luo, A. K. Y. Jen, and N. Peyghambarian, "Hybrid cross-linkable polymer/sol-gel waveguide modulators with 0.65 V half wave voltage at 1550 nm," *Applied Physics Letters,* vol. 91, p. 093505, 2007.

[22] J. Luo, S. Huang, Y. J. Cheng, T. D. Kim, Z. Shi, X. H. Zhou, and K. Y. J. Alex, "Phenyltetraene-based nonlinear optical chromophores with enhanced chemical stability and electrooptic activity," *Organic letters,* vol. 9, pp. 4471-4474, 2007.

[23] T. Baehr-Jones, M. Hochberg, G. Wang, R. Lawson, Y. Liao, P. Sullivan, L. Dalton, A. K. Y. Jen, and A. Scherer, "Optical modulation and detection in slotted silicon waveguides," *Optics Express,* vol. 13, pp. 5216-5226, 2005.

[24] J. H. Wülbern, J. Hampe, A. Petrov, M. Eich, J. Luo, A. K. Y. Jen, A. Di Falco, T. F. Krauss, and J. Bruns, "Electro-optic modulation in slotted resonant photonic crystal heterostructures," *Applied Physics Letters,* vol. 94, p. 241107, 2009.

[25] X. Wang, C. Y. Lin, S. Chakravarty, J. Luo, A. K. Y. Jen, and R. T. Chen, "Effective in-device $r_{33}$ of 735 pm/V on electro-optic polymer infiltrated silicon photonic crystal slot waveguides," *Optics letters,* vol. 36, pp. 882-884, 2011.

[26] A. C. G. Nutt, V. Gopalan, and M. C. Gupta, "Domain inversion in LiNbO3 using direct electron‐beam writing," *Applied Physics Letters,* vol. 60, pp. 2828-2830, 1992.

[27] D. Jin, H. Chen, A. Barklund, J. Mallari, G. Yu, E. Miller, and R. Dinu, "EO polymer modulators reliability study," in *Proc. SPIE 7599*, 2010, pp. 75990H-8.

[28] X. Wang, B. S. Lee, C. Y. Lin, D. An, and R. T. Chen, "Electroptic polymer linear modulators based on multiple-domain Y-fed directional coupler," *Journal of Lightwave Technology,* vol. 28, pp. 1670-1676, 2010.

[29] B. Lee, C. Lin, X. Wang, R. T. Chen, J. Luo, and A. K. Y. Jen, "Bias-free electro-optic polymer-based two-section Y-branch waveguide modulator with 22 dB linearity enhancement," *Optics letters,* vol. 34, pp. 3277-3279, 2009.

[30] FIMMWAVE simulation software, http://www.photond.com/products/fimmwave.htm

[31] FIMMPROP simulation software, http://www.photond.com/products/fimmprop.htm

[32] G. T. Reed, *Silicon Photonics: the state of the art*. UK: Wiley-Interscience, 2008.

[33] COMSOL Multiphysics Simulation Software, http://www.comsol.com.

[34] R. G. Hunsperger, A. Yariv, and A. Lee, "Parallel end-butt coupling for optical integrated circuits," *Applied optics,* vol. 16, pp. 1026-1032, 1977.

[35] M. Sanghadasa, P. R. Ashley, E. L. Webster, C. Cocke, G. A. Lindsay, and A. J. Guenthner, "A simplified technique for efficient fiber-polymer-waveguide power coupling using a customized cladding with tunable index of refraction," *Lightwave Technology, Journal of,* vol. 24, pp. 3816-3823, 2006.

[36] K. Geary, S. K. Kim, B. J. Seo, Y. C. Hung, W. Yuan, and H. R. Fetterman, "Photobleached refractive index tapers in electrooptic polymer rib waveguides," *Photonics Technology Letters, IEEE,* vol. 18, pp. 64-66, 2006.

[37] I. E. Day, I. Evans, A. Knights, F. Hopper, S. Roberts, J. Johnston, S. Day, J. Luff, H. K. Tsang, and M. Asghari, "Tapered silicon waveguides for low insertion loss highly-efficient high-speed electronic variable optical attenuators," in *Optical Fiber Communications Conference*, 2003, pp. 249 - 251

[38] Rsoft simulation software, http://www.rsoftdesign.com/

[39] W. S. Chang, *RF photonic technology in optical fiber links*: Cambridge Univ Pr, 2002.

[40] D. Chen, Q. Wang, and Z. Shen, "A broadband microstrip-to-CPW transition," in *Microwave Conference Proceedings, 2005. APMC 2005. Asia-Pacific Conference Proceedings*, 2005, p. 4.

[41] Y. G. Kim, K. W. Kim, and Y. K. Cho, "An ultra-wideband Microstrip-to-CPW transition," in *Microwave Symposium Digest, 2008 IEEE MTT-S International*, Atlanta, GA, 2008, pp. 1079-1082.

[42] E. Yamashita, "Variational method for the analysis of microstrip-like transmission lines," *Microwave Theory and Techniques, IEEE Transactions on,* vol. 16, pp. 529-535, 1968.

[43] ANSYS HFSS simulation software, http://www.ansys.com/

[44] A. Chen and E. Murphy, *Broadband Optical Modulators: Science, Technology, and Applications*: CRC Press, 2011.

[45] R. C. Alferness, "Waveguide electrooptic modulators," *IEEE Transactions on Microwave Theory Techniques,* vol. 30, pp. 1121-1137, 1982.

[46] D. M. Burland, R. D. Miller, and C. A. Walsh, "Second-order nonlinearity in poled-polymer systems," *Chemical Reviews,* vol. 94, pp. 31-75, 1994.

[47] L. R. Dalton, P. A. Sullivan, and D. H. Bale, "Electric field poled organic electro-optic materials: state of the art and future prospects," *Chemical Reviews,* vol. 110, pp. 25-55, 2009.

[48] M. A. Mortazavi, A. Knoesen, S. T. Kowel, B. G. Higgins, and A. Dienes, "Second-harmonic generation and absorption studies of polymer—dye films oriented by corona-onset poling at elevated temperatures," *JOSA B,* vol. 6, pp. 733-741, 1989.

[49] H. Tang, J. M. Taboada, G. Cao, L. Li, and R. T. Chen, "Enhanced electro-optic coefficient of nonlinear optical polymer using liquid contact poling," *Applied Physics Letters,* vol. 70, p. 538, 1997.

[50] Z. Z. Yue, D. An, R. T. Chen, and S. Tang, "1000 V/μm pulsed poling technique for photolime-gel electro-optic polymer with room-temperature repoling feature," *Applied Physics Letters,* vol. 72, p. 3420, 1998.

[51] S. Huang, T. D. Kim, J. Luo, S. K. Hau, Z. Shi, X. H. Zhou, H. L. Yip, and A. K. Y. Jen, "Highly efficient electro-optic polymers through improved poling using a thin TiO-modified transparent electrode," *Applied Physics Letters,* vol. 96, p. 243311, 2010.

[52] S. Huang, J. Luo, H. L. Yip, A. Ayazi, X. H. Zhou, M. Gould, A. Chen, T. Baehr‐Jones, M. Hochberg, and A. K. Y. Jen, "Electro‐optical Materials: Efficient Poling of Electro‐Optic Polymers in Thin Films and Silicon Slot Waveguides by Detachable Pyroelectric Crystals (Adv. Mater. 10/2012)," *Advanced Materials,* vol. 24, pp. OP1-OP1, 2012.

[53] R. Blum, M. Sprave, J. Sablotny, and M. Eich, "High-electric-field poling of nonlinear optical polymers," *JOSA B,* vol. 15, pp. 318-328, 1998.

[54] T. A. Tumolillo Jr and P. R. Ashley, "A novel pulse-poling technique for EO polymer waveguide devices using device electrode poling," *Photonics Technology Letters, IEEE,* vol. 4, pp. 142-145, 1992.

[55] V. Taggi, F. Michelotti, M. Bertolotti, G. Petrocco, V. Foglietti, A. Donval, E. Toussaere, and J. Zyss, "Domain inversion by pulse poling in polymer films," *Applied Physics Letters,* vol. 72, p. 2794, 1998.

[56] J. Baker-Jarvis, M. D. Janezic, B. Riddle, C. L. Holloway, and N. Paulter, "Dielectric and conductor-loss characterization and measurements on electronic packaging materials," *2001.,* 2001.

[57] G. K. Gopalakrishnan, W. K. Burns, R. W. McElhanon, C. H. Bulmer, and A. S. Greenblatt, "Performance and modeling of broadband LiNbO3 traveling wave optical intensity modulators," *Lightwave Technology, Journal of,* vol. 12, pp. 1807-1819, 1994.

[58] P. L. Liu, B. Li, and Y. Trisno, "In search of a linear electrooptic amplitude modulator," *Photonics Technology Letters, IEEE,* vol. 3, pp. 144-146, 1991.

[59] W. B. Bridges and J. H. Schaffner, "Distortion in linearized electrooptic modulators," *Microwave Theory and Techniques, IEEE Transactions on,* vol. 43, pp. 2184-2197, 1995.

[60] Y. C. Hung, S. K. Kim, H. Fetterman, J. Luo, and A. K. Y. Jen, "Experimental demonstration of a linearized polymeric directional coupler modulator," *Photonics Technology Letters, IEEE,* vol. 19, pp. 1762-1764, 2007.

[61] C. T. DeRose, R. Himmelhuber, D. Mathine, R. Norwood, J. Luo, A. K. Y. Jen, and N. Peyghambarian, "High Δn strip-loaded electro-optic polymer waveguide modulator with low insertion loss," *Optics Express,* vol. 17, pp. 3316-3321, 2009.